\documentclass{PoS}

\title{{\sc Cristal} and {\sc Azurite}: new tools for integration-by-parts reductions}

\ShortTitle{{\sc Cristal} and {\sc Azurite}: new tools for integration-by-parts reductions}

\author{Alessandro Georgoudis\\
        Department of Physics and Astronomy, Uppsala University, SE-75108 Uppsala, Sweden\\
        E-mail: \email{Alessandro.Georgoudis@physics.uu.se}}

\author{\speaker{Kasper J.~Larsen}\\
       School of Physics and Astronomy, University of Southampton, Highfield, Southampton, \\
       SO17 1BJ, United Kingdom\\
       E-mail: \email{Kasper.Larsen@soton.ac.uk}}

\author{Yang Zhang\\
        PRISMA Cluster of Excellence, Johannes Gutenberg University, 55128 Mainz, Germany\\
        E-mail: \email{zhang@uni-mainz.de}}

\abstract{
Scattering amplitudes computed at a fixed loop order,
along with any other object computed in perturbative
quantum field theory, can be expressed as a
linear combination of a finite basis of loop integrals.
To compute loop amplitudes in practice, such a basis of integrals must be determined.
We discuss {\sc Azurite} ({\bf A ZUR}ich-bred method for
finding master {\bf I}n{\bf TE}grals), a publicly available
package for finding bases of loop integrals.
We also discuss {\sc Cristal} ({\bf C}omplete {\bf R}eduction of {\bf I}ntegral{\bf S}
{\bf T}hrough {\bf A}ll {\bf L}oops), a future package that produces the complete
integration-by-parts reductions.
}

\FullConference{13th International Symposium on Radiative Corrections (Applications of Quantum Field Theory to Phenomenology)\\
         25-29 September, 2017 \\
         St. Gilgen, Austria}

\def\d{{\rm d}}
\usepackage{tikz}
\usepackage{amsmath}
\usepackage{amssymb}
\usepackage[numbers, sort&compress]{natbib}

\begin{document}

\section{Introduction}

Precision calculations of the cross sections of Standard Model
processes at the Large Hadron Collider are crucial to gain a
quantitative understanding of the background and in turn improve the
ability to extract signals of new physics. This typically requires
computations at next-to-next-to leading order (NNLO) in fixed-order
perturbation theory, in order to match the experimental precision and
the parton distribution function uncertainties.

A key tool in these calculations are integration-by-parts (IBP)
identities \cite{Tkachov:1981wb, Chetyrkin:1981qh,Laporta:2000dc,%
Laporta:2001dd,Smirnov:2003kc,Anastasiou:2004vj,Smirnov:2008iw,Studerus:2009ye,%
Gluza:2010ws,vonManteuffel:2012yz,Lee:2012cn,vonManteuffel:2014ixa,%
Smirnov:2014hma,Ita:2015tya,Larsen:2015ped}. These are relations that arise from the vanishing
integration of total derivatives,
\begin{equation}
  \int \prod_{j=1}^L \bigg (\frac{\d^D \ell_j}{\mathrm{i} \pi^{D/2}}\bigg)
  \sum_{i=1}^L \frac{\partial}{\partial \ell^\mu_i}
  \frac{v_i^\mu}{D_1^{\alpha_1} \cdots D_k^{\alpha_k}}=0\,,
  \label{eq:IBP_schematic}
\end{equation}
where the vectors $v_i^\mu$ are polynomials in the internal and
external momenta, the $D_k$ denote inverse propagators, and the $\alpha_i \geq 1$ are
integers. The IBP identities allow the loop integrals that contribute
to a loop-level quantity, say a scattering amplitude, to be expressed
in a finite basis of integrals.%
\footnote{The fact that the basis of integrals is always finite was
proven in ref.~\cite{Smirnov:2010hn}.}
In practice, this leads to
a very significant simplification of the representation of the amplitude.
IBP reductions moreover allow setting up differential equations
for the basis integrals, thereby enabling their evaluation
\cite{Kotikov:1990kg,Kotikov:1991pm,Bern:1993kr,Remiddi:1997ny,Gehrmann:1999as,Henn:2013pwa}.

An initial step of generating IBP reductions is to determine a basis of integrals.
In this proceedings contribution we discuss the {\sc Singular} \cite{DGPS}/{\sc Mathematica}
package {\sc Azurite} ({\bf A ZUR}ich-bred method for finding master
{\bf I}n{\bf TE}grals) \cite{Georgoudis:2016wff} which determines a basis for the space of integrals
spanned by a given $L$-loop diagram and all of its subdiagrams
(obtained by pinching lines). As the underlying algorithm is
completely general, {\sc Azurite} can be used to provide a basis
for any number of loops and external particles, arbitrary configurations
of internal and external masses in both the planar and non-planar sectors.
In practice, the running time for two-loop diagrams is typically
of the order of a few seconds, and for three-loop diagrams a few minutes at most.
{\sc Azurite} has been tested up to five loops.

Related work has appeared in ref.~\cite{Lee:2013hzt} where the number
of basis integrals is determined as the sum of the Milnor invariant
evaluated at the critical points of the polynomials that
enter the parametric representation. This method has moreover
been implemented in the {\sc Mathematica} package {\tt Mint}.

\section{Integration-by-parts identities on maximal cuts}\label{sec:IBPs_on_maximal_cuts}

We start by setting up notation and conventions.
We consider an $L$-loop integral with $k$ propagators and
$m-k$ irreducible scalar products (i.e., polynomials in the
loop momenta and external momenta which cannot be expressed
as a linear combination of the inverse propagators).
We work in dimensional regularization and normalize the
integral as follows,
\begin{equation}
I(\alpha_1, \ldots, \alpha_m; D) \equiv \int \prod_{j=1}^L \frac{\d^D \ell_j}{\mathrm{i} \pi^{D/2}}
\frac{D_{k+1}^{\alpha_{k+1}} \cdots D_m^{\alpha_m}}{D_1^{\alpha_1} \cdots D_k^{\alpha_k}}
\hspace{4mm} \mathrm{with} \hspace{4mm} \alpha_i \geq 0 \,.
\label{eq:def_generic_Feynman_integral}
\end{equation}

{\sc Azurite} determines integration-by-parts identities (\ref{eq:IBP_schematic})
on unitarity cuts, where some set of propagators are put on shell,
$D_i = 0$ for $i \in \mathcal{S}$. To this end we make use of
the Baikov representation \cite{Baikov:1996rk} whose variables are
the inverse propagators and the irreducible scalar products,
$z_\alpha \equiv D_\alpha$ where $1 \leq \alpha \leq m$.
In order to write down the associated Jacobian, we define the
set of independent external and loop momenta,
$\{ v_1, \ldots, v_{n+L-1} \}$ $= \{ p_1, \ldots, p_{n-1}, \ell_1, \ldots, \ell_L \}$
and consider the associated Gram determinant,
$F= \det_{i,j=1, \hspace{0.5mm} \ldots, \hspace{0.5mm} n+L-1} (v_i \cdot v_j)$.
We can now express the integral in eq.~(\ref{eq:def_generic_Feynman_integral})
in its Baikov representation,
\begin{equation}
I(\alpha; D) \hspace{0.5mm}\propto\hspace{0.5mm}
\int \d z_1 \cdots \d z_m \frac{z_{k+1}^{\alpha_{k+1}} \cdots z_m^{\alpha_m}}{z_1^{\alpha_1} \cdots z_k^{\alpha_k}} F(z)^\frac{D-L-n}{2} \,.
\label{eq:Baikov_representation}
\end{equation}
To find integration-by-parts identities on the maximal cut
$z_1 = \cdots = z_k = 0$ of this diagram, we consider
the most general total derivative of the same form as the residue
of eq.~(\ref{eq:Baikov_representation}) at this pole,
\begin{align}
0 &= \int \frac{\d z_{k+1} \cdots \d z_m}{z_{k+1}^{-\alpha_{k+1}} \cdots z_m^{-\alpha_m}}
\sum_{i=k+1}^m \frac{\partial}{\partial z_i} \Big( a_i(z) F(z)^\frac{D-n-L}{2} \Big)
\label{eq:total_derivative_on_max_cut_1}\\
&= \int \frac{\d z_{k+1} \cdots \d z_m}{z_{k+1}^{-\alpha_{k+1}} \cdots z_m^{-\alpha_m}}
\sum_{i=k+1}^m \left( \frac{\partial a_i}{\partial z_i}
+ \frac{D-n-L}{2 F(z)} a_i(z) \frac{\partial F}{\partial z_i} \right) F(z)^\frac{D-n-L}{2} \,,
\label{eq:total_derivative_on_max_cut_2}
\end{align}
where the $a_i(z)$ denote polynomials. We observe that,
for an arbitrary choice of $a_i(z)$, the terms in the parenthesis $(\cdots)$
in eq.~(\ref{eq:total_derivative_on_max_cut_2}) correspond to
integrals in $D$ and $D-2$ dimensions, respectively.
This is because the $\frac{1}{F(z)}$ factor in the
second term effectively modifies the integration measure,
shifting the space-time dimension from $D$ to $D-2$.

We can avoid the dimension shift by choosing the $a_i(z)$ such
that,
\begin{equation}
\sum_{i=k+1}^m a_i (z) \frac{\partial F}{\partial z_i} + b F = 0 \,,
\label{eq:syzygy_equation}
\end{equation}
where $b$ denotes a polynomial, since then
the $\frac{1}{F}$ factor in eq.~(\ref{eq:total_derivative_on_max_cut_2})
cancels out, and the ansatz in eq.~(\ref{eq:total_derivative_on_max_cut_1})
corresponds to an integration-by-parts identity in purely $D$ dimensions.

Equations of the form (\ref{eq:syzygy_equation}) are known in algebraic
geometry as \emph{syzygy equations}. Algorithms for obtaining a
generating set of solutions are known and have been implemented
in several computer algebra systems dedicated to computational algebraic geometry,
such as Singular \cite{DGPS} and Macaulay2 \cite{M2}.
By plugging the obtained set of solutions of eq.~(\ref{eq:syzygy_equation})
into eq.~(\ref{eq:total_derivative_on_max_cut_2}), we then find
the required integration-by-parts identities evaluated on the maximal cut $z_1 = \cdots = z_k = 0$.

\section{Algorithm of {\sc Azurite}}

The algorithm of {\sc Azurite} can be summarized as follows.
Given an input set of inverse propagators $D_1, \ldots, D_k$
and irreducible numerator insertions $D_{k+1}, \ldots, D_m$,
{\sc Azurite} determines a basis of integrals by proceeding
through the following steps.
\begin{enumerate}
\item Find the corresponding graph $\Gamma$ and the automorphism groups of $\Gamma$ and all of its subgraphs.

\item Find a list $\mathcal{C}$ of cuts such that
no two elements of $\mathcal{C}$ are related by any of the symmetries
found in step 1.

\item For each cut $c \in \mathcal{C}$, construct IBP identities
and symmetry relations on $c$, giving $\mathbb{Z}_p$ values
to external kinematical invariants
and the space-time dimension $D$ for efficiency.

\item For each cut $c \in \mathcal{C}$, apply Gauss-Jordan elimination
to the system of symmetry relations and IBP identities found in step 3.
The resulting non-pivot entries correspond to basis integrals.
\end{enumerate}

In the current implementation of {\sc Azurite}, step 1 is realized
by graph theory functions in {\sc Mathematica}. The construction of
IBP identities on maximal cuts in step 3 is carried out with the
formalism explained in section \ref{sec:IBPs_on_maximal_cuts}.
In setting up the linear systems to which Gauss-Jordan elimination
is applied in step 4, {\sc Azurite} sorts the integrals
in descending order by the following order relation. For two integrals
$I(\alpha; D)$ and $I(\beta; D)$ with propagator powers
$\alpha = (\alpha_1, \ldots, \alpha_k)$ and $\beta = (\beta_1, \ldots, \beta_k)$,
we have $I(\alpha; D) > I(\beta; D)$ if $\sum_{i=1}^k |\alpha_i| > \sum_{i=1}^k |\beta_i|$
or, in the event of a tie, $\sum_{i=k+1}^m |\alpha_i| > \sum_{i=k+1}^m |\beta_i|$
or, in the event of a further tie, $(\alpha_1, \ldots, \alpha_m) > (\beta_1, \ldots, \beta_m)$
lexicographically.
To improve efficiency, {\sc Azurite} makes use of two graph-based
simplifications: 1) diagrams with massless tadpoles are detected and
discarded; and 2) on any given cut $c$, rather than
using the parent momenta of the top-level graph, {\sc Azurite} uses the
available momenta (which may be sums of several parent momenta). This
reduces the number of scalar products.

\section{Performance of {\sc Azurite}}

The figure below shows the computation time (on a single core on a standard laptop)
and number of basis integrals for a variety of diagrams at various loop orders
and configurations of internal and external masses.

\begin{figure}[!h]
\begin{center}
    \includegraphics[width=0.78\textwidth]{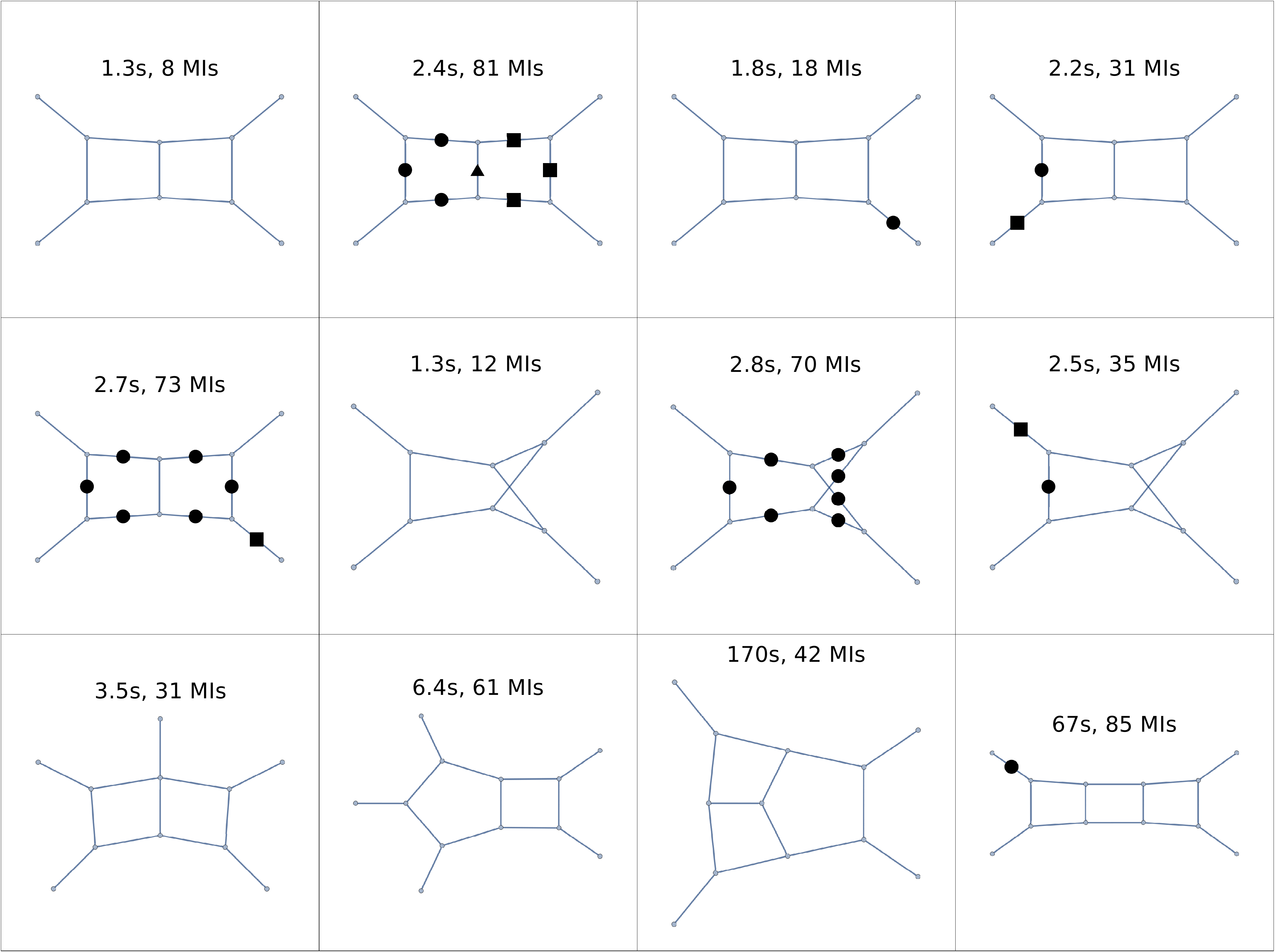}
\end{center}
\caption{Computation time and number of basis integrals for different topologies and mass configurations.
Here, $\blacktriangle$, $\blacksquare$ and $\bullet$ represent different masses.}
\label{fig:TimingAndResults}
\end{figure}

\clearpage

\begin{figure}[!h]
\begin{center}
    \includegraphics[width=0.4\textwidth]{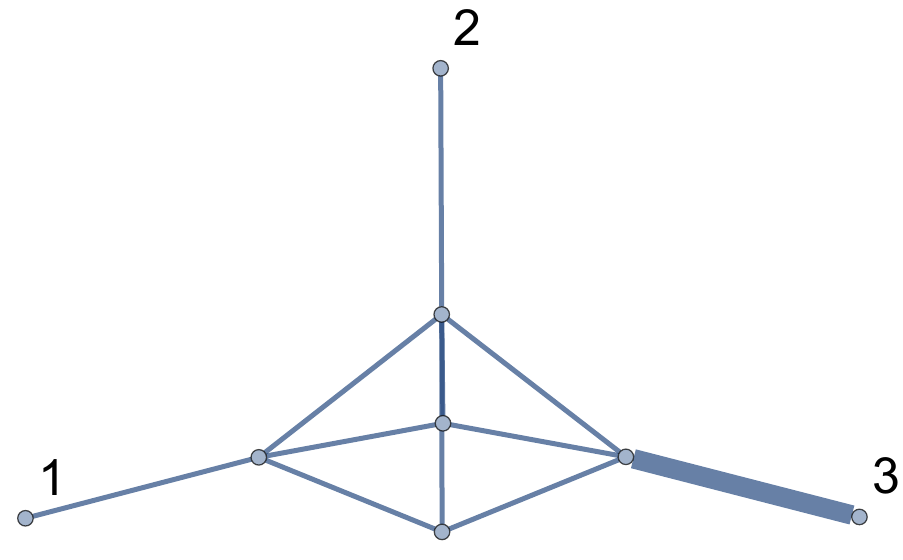}
\end{center}
\caption{Integral topology for the four-loop form factor.
For this diagram, the computation time to determine an integral basis
is reduced from 180 s in version 1 of {\sc Azurite} to 41 s in version 2.}
\label{fig:four-loop_form_factor}
\end{figure}

In a new version, {\sc Azurite} 2 \cite{Georgoudis:Azurite2},
to be made publicly available shortly, the computation time
is reduced significantly, particularly
for three- and higher-loop topologies.
An example is provided in fig.~\ref{fig:four-loop_form_factor}.

\section{{\sc Cristal}}

{\sc Cristal} ({\bf C}omplete {\bf R}eduction of {\bf I}ntegral{\bf S}
{\bf T}hrough {\bf A}ll {\bf L}oops) is a future package that will produce the complete
integration-by-parts reductions. It is based on the formalism developed in
ref.~\cite{Larsen:2015ped}. The basic idea is to study the integration-by-parts
identities (\ref{eq:IBP_schematic}) on a spanning set of cuts,
defined as the maximal cuts of those elements of an integral basis $\mathcal{B}$
which cannot be obtained from another $b\in \mathcal{B}$ by adding propagators,
\begin{equation}
\mathcal{C} = \{ c \in \mathcal{B}: \nexists b \in \mathcal{B}: \hspace{1.5mm} b \mbox{ is a strict subgraph of } c \} \,.
\label{eq:spanning_set_of_cuts}
\end{equation}
The effect of constructing the IBP reductions on the set of cuts $\mathcal{C}$
is, roughly speaking, to block-diagonalize the linear systems to which Gauss-Jordan
elimination is applied in the standard Laporta algorithm.
We emphasize that a spanning set of cuts (\ref{eq:spanning_set_of_cuts}) can
only be determined once an integral basis $\mathcal{B}$ is known. Thus,
{\sc Azurite} provides an initial step for the full reduction problem
undertaken by {\sc Cristal}.

\begin{figure}[!h]
\begin{center}
    \includegraphics[width=0.8\textwidth]{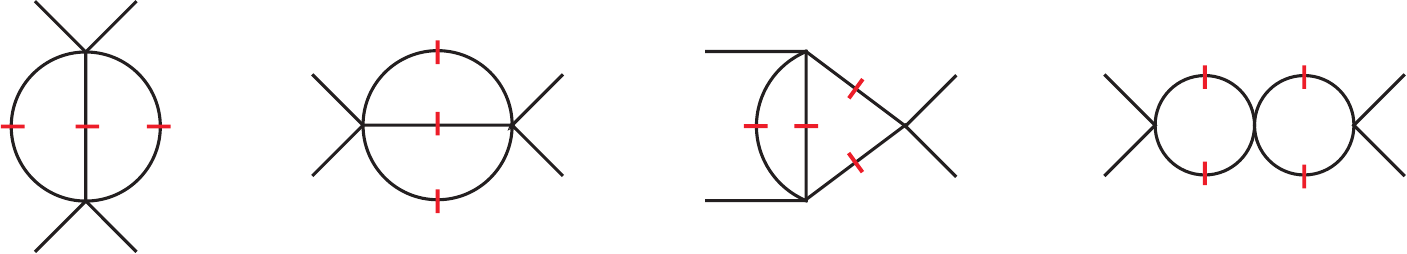}
\end{center}
\caption{Complete IBP reductions can be efficiently obtained by
studying the IBP identities on a spanning set of cuts, defined in eq.~(\ref{eq:spanning_set_of_cuts}).}
\label{fig:SpanningSetOfCuts}
\end{figure}

\clearpage

\bibliography{CristalAzurite}


\end{document}